\documentclass[conference]{IEEEtran}

\IEEEoverridecommandlockouts
\usepackage{verbatim}
\usepackage{longtable}
\DeclareMathAlphabet{\mathpzc}{OT1}{pzc}{m}{it}
\usepackage{amsmath}
\usepackage{amsfonts}
\usepackage{amssymb}
\usepackage{multirow}
\usepackage{amsmath,amssymb,amsfonts}
\usepackage{fancyhdr}
\usepackage[ruled,vlined,linesnumbered]{algorithm2e}
\usepackage[
top    = 0.7in,
bottom = 0.95in,
left   = 0.65in,
right  = 0.57in]{geometry}
\usepackage{graphicx}
\usepackage{textcomp}
\usepackage{xcolor}
\usepackage{subcaption}
\usepackage{algpseudocode}
\usepackage{enumitem}
\usepackage{float}
\restylefloat{figure}
\usepackage{graphicx}
\usepackage[normalem]{ulem}
\useunder{\uline}{\ul}{}

\begin{document}

\title{Maximizing Blockchain Performance: Mitigating Conflicting Transactions through Parallelism and Dependency Management}

\author{\IEEEauthorblockN{Faisal Haque Bappy$^{1}$, Tarannum Shaila Zaman$^{2}$, Md Sajidul Islam Sajid$^{3}$,\\Mir Mehedi Ahsan Pritom$^{4}$, and Tariqul Islam$^{5}$}
\IEEEauthorblockA{
$^{1, 5}$ School of Information Studies (iSchool), Syracuse University, Syracuse, NY, USA\\
$ ^{2}$ Computer and Information Science, SUNY Polytechnic Institute, NY, USA\\
$ ^{3}$ Computer and Information Sciences, Towson University, Towson, MD, USA\\
$ ^{4}$ Computer Science, Tennessee Tech University, Cookeville, TN, USA\\
Email: \{fbappy@syr, zamant@sunypoly, msajid@towson, mpritom@tntech, mtislam@syr\}.edu} 
}

\maketitle

\thispagestyle{fancy}
 \lhead{This work has been accepted at the 7th IEEE International Conference on Blockchain (Blockchain 2024)}
\cfoot{}

\begin{abstract}

While blockchains initially gained popularity in the realm of cryptocurrencies, their widespread adoption is expanding beyond conventional applications, driven by the imperative need for enhanced data security. Despite providing a secure network, blockchains come with certain tradeoffs, including high latency, lower throughput, and an increased number of transaction failures. A pivotal issue contributing to these challenges is the improper management of ``conflicting transactions," commonly referred to as ``contention". When a number of pending transactions within a blockchain collide with each other, this results in a state of contention. This situation worsens network latency, leads to the wastage of system resources, and ultimately contributes to reduced throughput and higher transaction failures. In response to this issue, in this work, we present a novel blockchain scheme that integrates transaction parallelism and an intelligent dependency manager aiming to reduce the occurrence of conflicting transactions within blockchain networks. In terms of effectiveness and efficiency, experimental results show that our scheme not only mitigates the challenges posed by conflicting transactions, but also outperforms both existing parallel and non-parallel Hyperledger Fabric blockchain networks achieving higher transaction success rate, throughput, and latency. The integration of our scheme with Hyperledger Fabric appears to be a promising solution for improving the overall performance and stability of blockchain networks in real-world applications.

\end{abstract}

\begin{IEEEkeywords}
Blockchain, Conflicting Transaction, Transaction Parallelism, Dependency Management, Throughput, Hyperledger Fabric
\end{IEEEkeywords}

\section{Introduction}
Blockchain technology, renowned for its decentralized and secure nature, has attracted considerable attention across various industries. Hyperledger Fabric \cite{HyperledgerFabric}, a prominent permissioned blockchain framework, is widely embraced for its modular architecture and enterprise-friendly features. However, as blockchain networks manage diverse and concurrent transactions, understanding the crucial factors that affect their performance becomes essential for ensuring efficient and reliable operations. 

Several works have concentrated on enhancing the performance of blockchain technology. The primary objective behind improving performance is to render blockchain usable and easily scalable, enabling its integration into more traditional systems that require an additional layer of security. Some efforts have focused on enhancing performance by customizing the consensus protocol to process transactions more efficiently \cite{khan2018fast}. However, these consensus algorithms have limitations and are only applicable to specific use cases.

Other researchers propose configuring the orderer structure to expedite the processing. FastFabric \cite{gorenflo2020fastfabric} and XOXFabric \cite{gorenflo2020xox} have demonstrated significant performance improvements by adjusting the ordering mechanisms of the default Hyperledger Fabric. Some authors have also advocated for the incorporation of parallel ordering to enhance performance further \cite{amiri2019parblockchain}. 

Even with the proposed solutions, the blockchain system faces challenges when dealing with conflicting transactions, causing a significant number of transactions to fail. This means that, even if networks can process more transactions in a second, they cannot ensure a higher success rate. Simply increasing throughput, or the number of transactions processed, is not very effective unless it translates into a better transaction success rate. In the end, the goal is to improve the network's overall effectiveness for users by ensuring that a higher proportion of transactions successfully go through despite potential conflicts.

In this paper, we focus on contention issues in blockchain arising from conflicting transactions. The main objective is to demonstrate how effectively we can manage conflicting transactions and thereby enhance the performance of blockchain without any modification to the orderers or consensus algorithms. For benchmarking purposes, we compare the performance of our scheme with the fastest version of Fabric (i.e.,  FastFabric), and the default Fabric. The following are the major contributions of this paper.

\begin{itemize}
\item {\bf Formal Definition of Contention in Blockchain Ordering Paradigms:} We clearly outlined and described the concept of contention within various ordering methods used in blockchain networks. This involves specifying how conflicts arise and impact the overall system.

\item {\bf Custom Contention Dataset Development:}
We created a custom contention dataset to rigorously test blockchain networks, focusing on scenarios involving a significant number of conflicting transactions. This dataset is crucial for assessing the robustness of the network under challenging conditions.

\item {\bf Simulation and Showcase the Impact of Contention on Hyperledger Fabric:} We conducted simulations to illustrate and highlight the consequences of contention on the performance of the default Hyperledger Fabric network. This involves demonstrating how contention affects the speed, reliability, and efficiency of the network.

\item {\bf Evaluation of Existing Solutions under Contentious Workloads:} We assessed and presented the effectiveness of current basic solutions in handling situations with intense contention. This involves evaluating how well existing approaches perform when faced with a high volume of conflicting transactions.

\item {\bf Performance Enhancement Approach and Demonstration:} We proposed a novel scheme to improve the overall performance of the blockchain network. This approach involves minimizing conflicts by implementing a dependency manager and utilizing parallel ordering, aiming to streamline and optimize transaction processing. Additionally, we demonstrated that our proposed approach outperforms the currently established schemes without necessitating additional modifications to the consensus and orderer structure of the blockchain network.
   
\end{itemize}

The remainder of the paper is structured as follows: in Section \ref{sec:prelim}, we describe some preliminary terms and concepts. In section \ref{sec:contention}, we explain the contention scenario in blockchain. Section \ref{sec:related} compares and contrasts existing works with our own. Section \ref{sec:simulation} presents how we simulate and analyze contention in the Hyperledger Fabric blockchain network. In Section \ref{sec:naive}, we demonstrate how existing naive solutions perform with highly contentious workloads. Then in Section \ref{sec:arch}, we present our proposed approach, followed by a performance analysis in Section \ref{sec:performance}. Finally, Section \ref{sec:conc} concludes the paper.

\section{Preliminaries}
\label{sec:prelim}

\subsection{Transaction Ordering}
Transaction ordering in blockchain determines the sequence of transactions in a block, affecting the state of the distributed ledger. Consensus mechanisms like Proof-of-Work (PoW) and Proof-of-Stake (PoS) are crucial in this process. In PoW, miners compete to solve mathematical problems; the first to solve one proposes a block with a specific transaction order \cite{gervais2016security}. PoS relies on participants with higher stakes to forge blocks \cite{larimer2013transactions}. Ensuring consistent transaction order across all nodes is vital for a consistent ledger state. Techniques like timestamping \cite{xu9317791} and deterministic algorithms \cite{peng2022neuchain} establish this order, enhancing blockchain reliability and security through consensus on transaction sequences.


\subsection{Contention}
In blockchain, contention occurs when multiple transactions simultaneously attempt conflicting operations on a single record, usually a previous block. Blockchain prioritizes fault tolerance and decentralized record-keeping without a central authority. State Machine Replication (SMR) synchronizes servers to ensure fault tolerance \cite{gorenflo2020xox}. However, malicious or faulty nodes can disrupt consensus \cite{xu9317791}. To counter this, Byzantine Fault Tolerant (BFT) protocols are used, though they face performance challenges with high contention workloads, where many transactions conflict \cite{amiri2019parblockchain}. Performance issues arise from the increased complexity of resolving conflicts and reaching consensus amid numerous contending transactions.

\section{Contention in Blockchain}
\label{sec:contention}
Ordering and Execution are the core parts of any blockchain system that ensures fault tolerance and distributed transaction processing. Based on the sequence of operations, there can be three types of paradigms. In this section, we will define the contention scenario in all these paradigms. 


\subsection{Order-Execute Permissionless:} 
\begin{figure}[]
    \centering
    \includegraphics[width=\columnwidth]{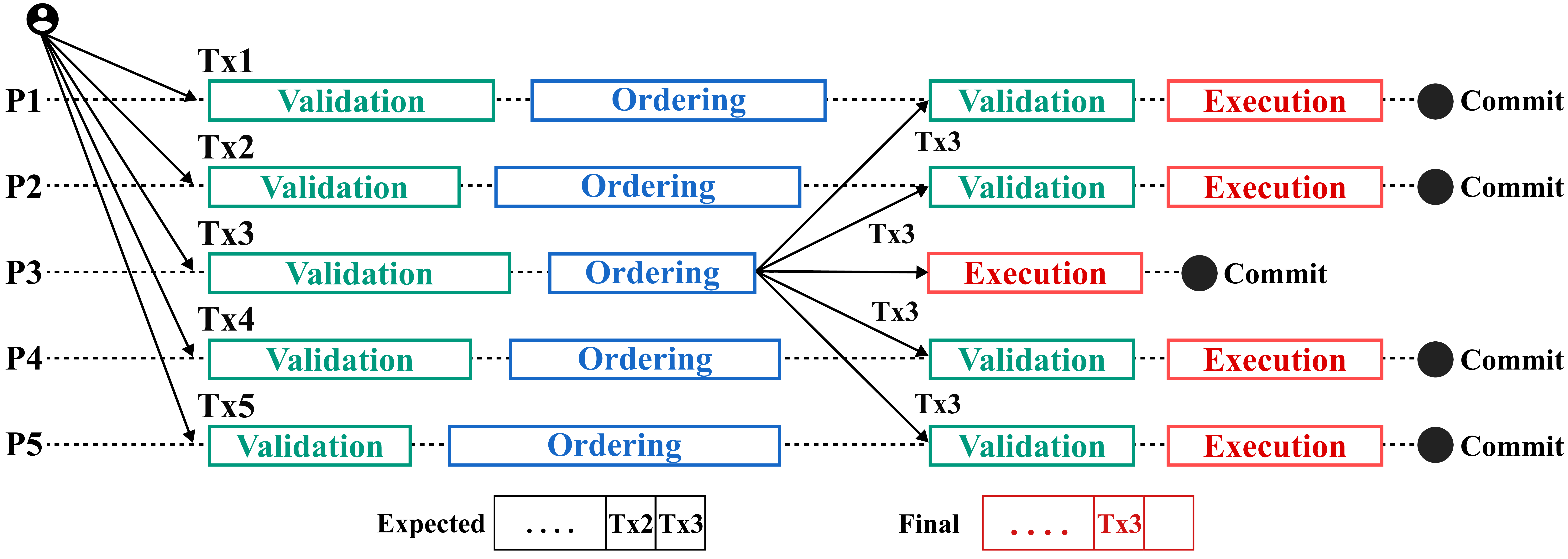}
            \caption{Contention in Order Execute Paradigm (Permissionless Blockchain)} \label{fig:ox-permissionless}
    \end{figure}

In this case, each transaction request from clients undergoes initial validation before peers initiate the ordering process. Upon successful ordering, the transaction is executed across all peers, and the new block is committed to the main chain. In the given scenario of Figure \ref{fig:ox-permissionless}, a client submitted five transaction requests: Tx1, Tx2, Tx3, Tx4, and Tx5. Notably, Tx3 has a dependency on Tx2, requiring Tx2 to be executed before Tx3 to maintain consistency.

However, with simultaneous requests, peers individually validate and order each transaction. If peer 3 orders Tx3 first and broadcasts it, other peers start execution, leading to inconsistency since Tx3 should follow Tx2. This may cause an execution error or invalidate the chain. Tx2 and Tx3 face contention due to their conflict. In large-scale networks, such conflicts are common, potentially causing delays and impeding transaction processing.

\subsection{Order-Execute Permissioned:} 
\begin{figure}[]
    \centering
    \includegraphics[width=\columnwidth]{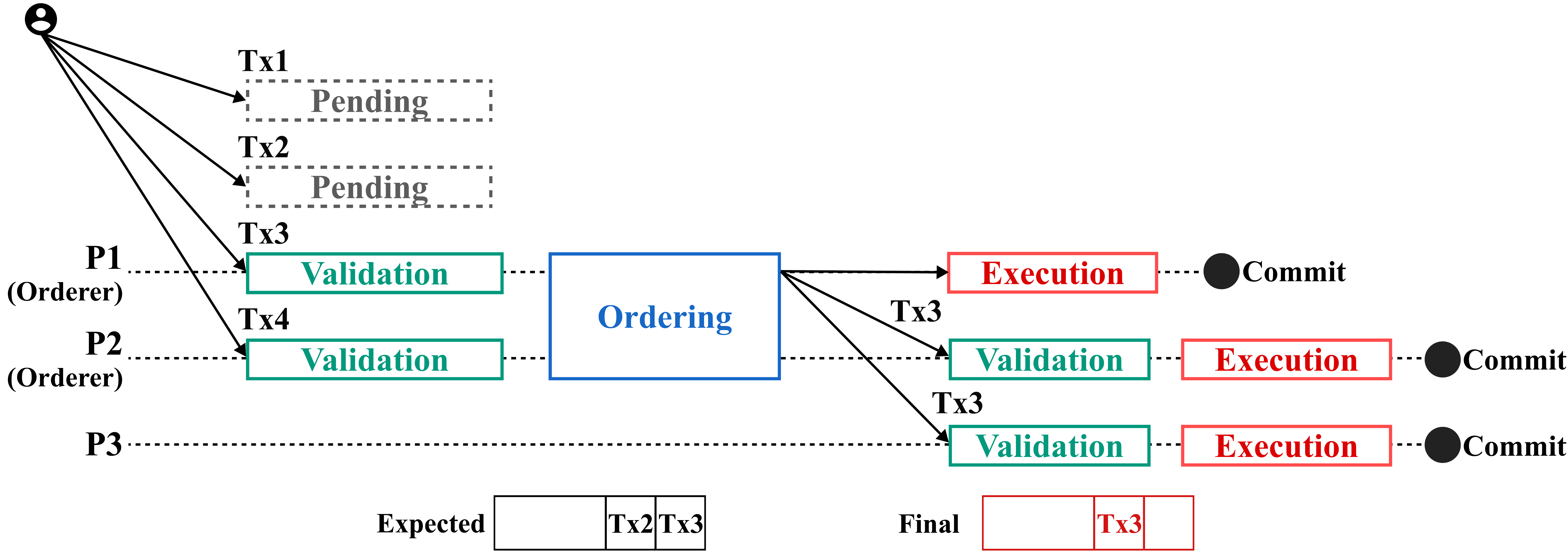}
            \caption{Contention in Order Execute Paradigm (Permissioned Blockchain)} \label{fig:ox-permissioned}
    \end{figure}

In permissioned blockchains, certain orderer nodes collaborate to validate and order transactions using a consensus protocol. In this particular scenario of Figure \ref{fig:ox-permissioned}, four transactions are submitted simultaneously, with only two orderer peers available. Consequently, the orderers initiate the ordering process for Tx3 and Tx4. Similar to the previous paradigm, Tx3 relies on Tx2, but Tx2 is queued, awaiting an available orderer.

If orderers P1 and P2 reach a consensus on Tx3, they broadcast it to other peers for execution. However, the inconsistency in ordering introduces the potential for execution errors or the creation of invalid blocks. Contentious transactions Tx2 and Tx3 encounter contention due to the unavailability of orderers. Many permissioned blockchains are designed for use within organizations with limited resources, making such cases likely to occur.
 
\subsection{Execute-Order Permissioned:} Certain permissioned blockchains, such as Hyperledger Fabric \cite{HyperledgerFabric}, employ a distinct paradigm known as Execute-Order. In this approach, requests are initially executed by trusted peers within the organization, referred to as endorsers. An endorsement policy, established by the network owner, dictates that clients must gather a specified number of endorsements before submitting for ordering.

\begin{figure}[]
    \centering
    \includegraphics[width=\columnwidth]{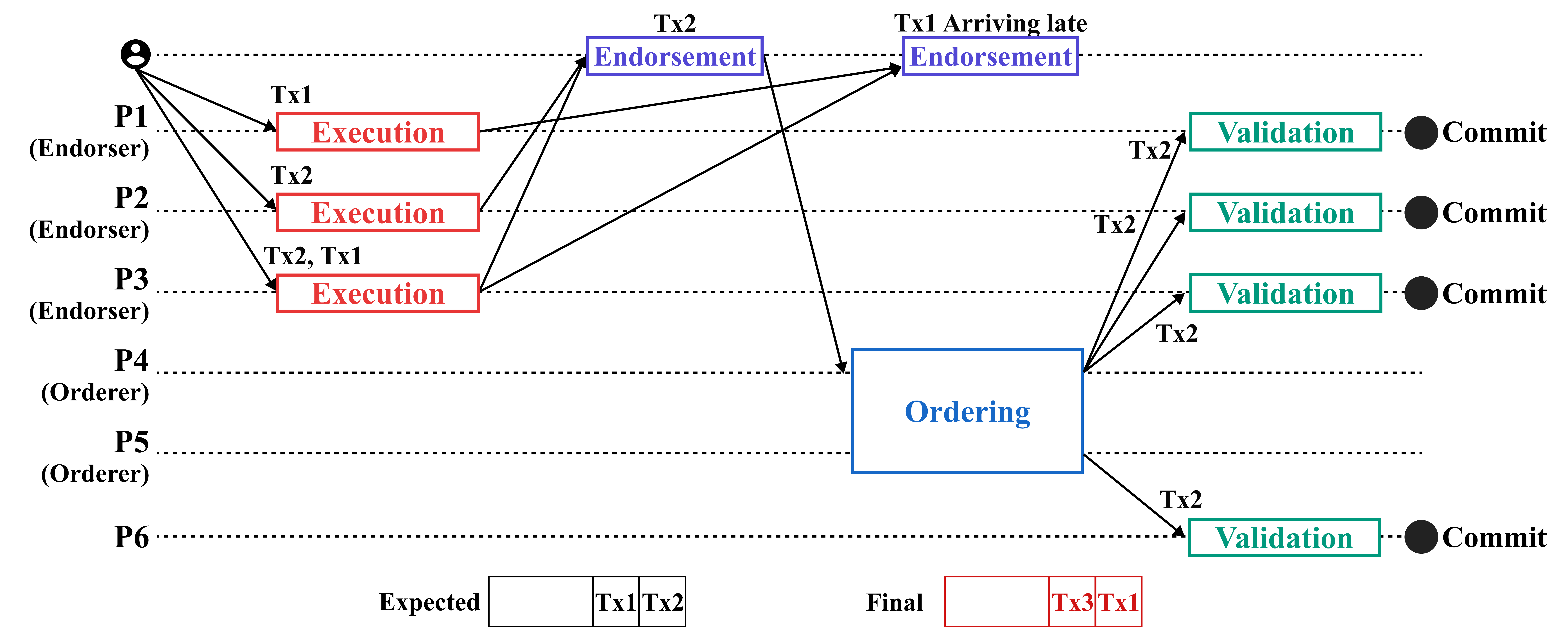}
            \caption{Contention in Execute Order Paradigm (Permissioned Blockchain)} \label{fig:xo-permissioned}
    \end{figure}

In this specific scenario of Figure \ref{fig:xo-permissioned}, two transactions, Tx1 and Tx2, exist, where Tx1 must be executed prior to Tx2. Endorsing peers P1, P2, and P3 are assigned, with P1 executing Tx1, P2 executing Tx2, and P3 executing both Tx1 and Tx2. Notably, Tx1 assigned to P3 will be executed only after P3 completes the execution of Tx2. If the client manages to collect the necessary endorsement for Tx2 before Tx1, it submits Tx2 for ordering, leading to inconsistency. As observed in the prior scenarios, this inconsistency results in an invalid block. Furthermore, since the transaction is executed before ordering, detecting the inconsistency from the execution order after ordering becomes challenging.

\section{Related Works}
\label{sec:related}
Due to the growing popularity of cryptocurrencies and other decentralized applications, blockchain systems are becoming increasingly complex and large-scale. This expansion results in a higher volume of transactions that need to be handled in a robust and secure manner. In response to this challenge, researchers have introduced the concept of concurrency through parallel processing. Amiri et al. introduced the ParBlockchain framework for the parallel processing of blockchain transactions, demonstrating how parallel processing can ensure faster transaction speed and high scalability in private blockchain networks \cite{amiri2019parblockchain}. 

However, as the processing of numerous transactions simultaneously increases, there is a corresponding rise in conflicting transactions, known as contentions. Contentions pose a prevalent problem in distributed systems. Salehi et al. conducted a comprehensive study on contention in distributed systems, creating a taxonomy for different types of contention and their respective countermeasures \cite{salehi2014taxonomy}. Various research works have proposed strategies to address contention based on specific scenarios, such as leasing resources, load balancing, location-aware mechanisms, and cache coherence \cite{salehi2014contention, kostin2000randomized, zhang2009location}. Kostin et al. proposed a load balancer-type approach for serving numerous client requests with an optimal number of servers; however, this approach does not guarantee robust performance for a large volume of requests, making it impractical for delay-sensitive applications \cite{kostin2000randomized}. Zhang et al. developed a framework to tackle contention using a location-aware caching mechanism for server allocation, addressing both performance and contention issues, but it is limited to centrally controlled systems \cite{zhang2009location}. Although these solutions were initially proposed for classical distributed systems, only a limited number of works have addressed contention issues in both private and public blockchain systems.

Nasirifard et al. \cite{fabricCRDT} and  Xu et al. \cite{xuLocking} converge on conflict resolution within Hyperledger Fabric. FabricCRDT proposed by Nasirifard et al., integrates Conflict-Free Replicated Datatypes to mitigate latency and transaction failures, while the Locking mechanism by Xu et al. optimizes performance, especially in transactions with concurrency conflicts. This linkage emphasizes the importance of conflict resolution strategies for maintaining the integrity of blockchain systems.

In terms of performance improvement, several authors have worked with different strategies before. Papers like FastFabric \cite{gorenflo2020fastfabric} and XOX Fabric \cite{gorenflo2020xox} converge on a shared goal of enhancing blockchain performance and scalability. FastFabric suggests tweaks to Hyperledger Fabric, significantly boosting transaction throughput. Simultaneously, XOX Fabric introduces a hybrid approach to transaction execution, alleviating performance bottlenecks during contention. Together, these papers lay the groundwork for blockchains that handle more transactions efficiently.

Beyond performance, several works focused on consensus protocols and fairness considerations. Kelkar et al. \cite{kelkar2020order} introduce a paradigm shift by emphasizing fairness in the order of transactions. Complementing this, Helix \cite{helix} proposes a consensus protocol resistant to ordering manipulation, ensuring fairness through randomized committee elections. These contributions lay the foundation for blockchain systems that not only reach consensus but do so fairly and tamper-resistant.

In the realm of permissioned blockchains, Goel et al. \cite{goel2018resource} innovatively tackle fair scheduling. The proposed weighted fair queuing strategy takes a nuanced approach to prioritize transactions based on their business importance. This connection introduces an essential element of fairness and efficiency into enterprise blockchain applications.

Some authors also focused on consensus mechanisms for performance improvement. Khan et al. \cite{khan2018fast} introduce FAST, a decentralized consensus mechanism leveraging the MapReduce paradigm. FAST addresses the throughput limitations of existing blockchain platforms, bringing them closer to the efficiency levels of traditional payment networks. This connection emphasizes the imperative of continually innovating consensus mechanisms for achieving high performance in blockchain systems.

While there has been substantial research focus on consensus mechanisms, fairness, and scalability within the realm of blockchain technology, a notable gap exists in the specific exploration of enhancing performance through conflict mitigation. This gap indicates an area that remains largely unexplored, representing an untapped potential for improving blockchain systems by addressing transactional conflicts.

\section{Simulating Contention}
\label{sec:simulation}
To gain a better understanding of how blockchain manages conflicts, we initiate the process by simulating conflicting transactions in a controlled environment. The aim is to observe how the orderer handles these conflicting transactions and identify potential points of failure. Previous research has suggested that certain schemes or frameworks perform better when dealing with conflicting transactions \cite{gorenflo2020xox, amiri2019parblockchain, gorenflo2020fastfabric}. However, the simulation environments used in these studies were not ideal for a comprehensive understanding of conflicts. Most studies relied on the SmallBank \cite{SmallBank} dataset for performance measurement, which primarily consists of regular transactions that may or may not result in conflicts. 

From our observations and initial experiments, we found that the number of conflicting transactions varies a lot in Hyperledger Fabric simulations. Sometimes, there were no conflicts at all \cite{benoit2021diablo}. So, we felt the need to create a custom testing dataset containing only conflicting transactions. When we simulate this dataset in high volume, it causes serious contention in blockchain networks. For our study, we made a dataset similar to SmallBank, but with only four wallets/accounts (A, B, C, D). It includes six types of transactions that can lead to conflicts when simulated together. \\

\textbf{Type 1: Transfer balance from A to both B and C}\\
In this transaction type, the objective is to simultaneously transfer a portion of the balance from wallet/account A to both wallet/accounts B and C. The potential for conflicts arises when multiple transactions attempt to transfer more balance than what is available in wallet A, leading to contention for the available funds.

\textbf{Type 2: Query the balance of C, while B is transferring assets to C}\\
This transaction involves querying the balance of wallet/account C while another transaction (Type 1) is in progress, transferring assets from B to C. Conflicts may arise when the balance query is executed at a point where the transfer from B to C is still ongoing, resulting in conflicting information about the balance.

\textbf{Type 3: Transfer assets from A, B, and C to D at the same time}\\
In this transaction type, assets are simultaneously transferred from wallets/accounts A, B, and C to wallet/account D. Conflicts may occur if the combined assets being transferred exceed the total available in A, B, and C, thereby testing the system's ability to handle concurrent transactions.

\textbf{Type 4: Transfer 100 random assets to A}\\
This transaction randomly transfers 100 assets to wallet/account A. Conflicts may arise if multiple transactions attempt to transfer assets to A concurrently, leading to contention for available assets and testing the system's resilience under such conditions.

\textbf{Type 5: Transfer 100 random assets from A}\\
Similar to Type 4, this transaction involves randomly transferring 100 assets from wallet/account A. Conflicts may occur if multiple transactions attempt to withdraw assets from A concurrently, challenging the system's ability to manage simultaneous asset transfers.

\textbf{Type 6: Transfer assets from A to D while it is unavailable}\\
This transaction type involves transferring assets from wallet/account A to D, with the added complexity that D is temporarily unavailable. Conflicts may arise when transactions attempt to transfer assets to D while it is inaccessible, testing the system's response and contention resolution when dealing with temporarily unavailable accounts.

When we simulated the dataset using the Hyperledger Caliper \cite{HyperledgerCaliper} tool on the Hyperledger Fabric blockchain, we observed a significant impact on performance in terms of throughput and latency. This highlights the substantial influence of contention on the overall performance of the blockchain. Our simulation involved using Hyperledger Caliper to generate 1000 transactions, specifically focusing on these six types, within a Hyperledger Fabric network. The network comprised four nodes running on Ubuntu 22.04 VMs, each with 8GB RAM and equipped with 4-core Standard D4s v3 CPUs from Azure. These CPUs typically operate at a base clock speed of around 2.3 GHz per core, making them suitable for moderate to high-performance computing tasks. 

To measure the impact, we compared the performance results with Fabric's default FabCar chaincode. The findings revealed that contentious workloads had a drastic effect (Table \ref{tab:sim-result}), leading to a considerable reduction in the overall performance of Hyperledger Fabric. \\

\begin{table}
\centering
\caption{Simulation Result from the Contention Workload}
\label{tab:sim-result}
\resizebox{0.8\columnwidth}{!}{
\begin{tabular}{l*{4}{c}}
\bf{TX Type}   & \bf{Success} & \bf{Fail} & \bf{TPS} & \bf{Latency(sec)} \\ 
 \hline \vspace{1mm}
\textit{Read} 		       & 1000 & 0 & 400 & 0.4 \\
 \vspace{1mm}
\textit{Write}            	& 987 & 13 & 70 & 4.6 \\
\vspace{1mm}
\textit{Update}           	& 873 & 127 & 43 & 6.5 \\ 
\hline \hline  \vspace{1mm} 
\textit{Type 1} 		    & 520 & 480 & 20 & 11.4 \\ 
\vspace{1mm}
\textit{Type 2}            	& 410 & 590 & 12 & 23.1 \\
\vspace{1mm}
\textit{Type 3}            	& 689 & 311 & 13.5 & 21.7 \\
\vspace{1mm}
\textit{Type 4}            	& 500 & 500 & 19.2 & 18.9 \\
\vspace{1mm}
\textit{Type 5}            	& 763 & 237 & 21.3 & 22.1 \\
\vspace{1mm}
\textit{Type 6}           	& 674 & 326 & 17.09 & 17.8 \\ 
\hline   
\end{tabular}
}
\end{table}

\section{Experiment with Naive Solutions}
\label{sec:naive}
As we have shown in our simulation, a contentious workload is one of the main reasons for hampering blockchain performance. To reduce contention in a blockchain network, it is crucial to handle the dependency of the transactions appropriately rather than just focusing on faster computation. There are some existing approaches for tackling contention, such as locking \cite{xuLocking}, grouping by transaction type \cite{fabricCRDT}, and timestamping \cite{xu9317791}. All of these approaches have demonstrated improvements in handling conflicting transactions. Yet, we aimed to delve further into their impact on blockchain performance and therefore tested three naive solutions individually and compared their results to gain deeper insights.

\subsection{Timestamping Approach}
In this approach, when a transaction is proposed, it is assigned a timestamp that reflects the time at which it was initiated. This timestamp can then be used to determine the order of transactions. This method significantly decreases the number of failed jobs. We applied this solution to the SmallBank dataset \cite{SmallBank} and utilized the Hyperledger Caliper \cite{HyperledgerCaliper} benchmark tool for performance measurement. Nevertheless, drawbacks accompany this solution, particularly latency. The process of synchronizing the timestamp database among all peers is time-consuming. Moreover, the orderers solely prioritize the timestamp, leading to potentially time-intensive transactions and subsequently high latency.

\subsection{Grouping Approach}
This approach focuses on categorizing transactions based on their type, such as read or write operations, user-defined priority, and estimated resource consumption. The ordering process mirrors the timestamping approach, with the orderer examining the transaction type and processing it according to the defined policy. In our implementation, we prioritized read operations over write and update operations, assuming a read-heavy system. However, this is a flexible choice for the developers to configure the priority depending on specific use cases. While this solution effectively reduces the job failure rate, it exhibits slightly higher rates compared to the timestamping method. However, it boasts the advantage of low latency, demonstrating performance akin to the default fabric in terms of latency.

\subsection{Locking Approach}
In this approach, We aimed to create a system similar to shared locking in the ordering process of Hyperledger Fabric. We set up a table to monitor access to wallets used in fabric smart contracts. This allowed the orderer to keep track of which wallets are currently in use and which ones are free. When a new endorsed transaction happens, the orderer checks the associated wallets. If everything needed is available, the transaction is carried out right away. If not, it goes into a queue and waits until the necessary resources are free. Although this locking system is effective in reducing job failures, with only a few issues related to incorrect addresses and values, further optimization is necessary to ensure completely conflict-free ordering. The downside, however, is that it takes a considerable amount of time, resulting in the longest delays compared to the other solutions we considered.

\begin{figure}[]
\centering
\includegraphics[width=\columnwidth]{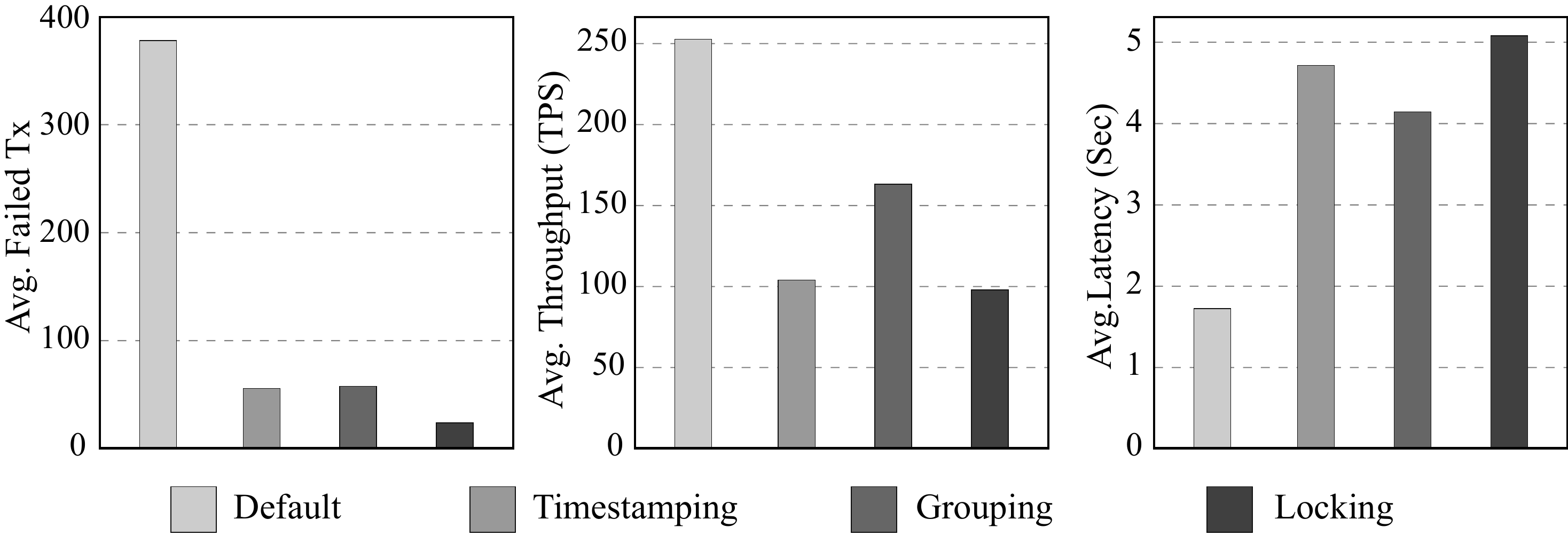}
        \caption{Performance of Naive Approaches} \label{fig:naive}
\end{figure}

\section{System Architecture}
\label{sec:arch}
In our proposed architecture, we have modified the existing Hyperledger Fabric framework to implement our solution and conduct all experiments and simulations. We will refer to our proposed architecture as ConChain. At a high level, ConChain comprises two main components integrated with the existing components of Hyperledger Fabric: the dependency manager and the transaction assigner, which manages all parallel processing tasks. In this section, we will elaborate on the structure and workflow of these two main components and how they handle conflicting transactions.
\subsection{Dependency Manager}
The Dependency Manager plays a crucial role in determining the dependent wallets of a transaction within ConChain. By default, all transactions in Hyperledger Fabric contain wallet addresses. However, there is currently no mechanism to detect other dependent wallets from these addresses. In ConChain, we address this gap by incorporating the dependency manager as a module connected to the endorsers.

Upon endorsement of a transaction, the dependency manager examines the addresses listed within that transaction. Initially, it checks whether each address is associated with a wallet or not. Subsequently, the dependency manager categorizes the wallets based on their access type. If the transaction requires read access to a wallet, it is included in the $readWallet$ list; otherwise, it is placed in the $writeWallet$ list. Additionally, the dependency manager verifies the validity of each wallet. If a wallet is found to be invalid, the transaction is discarded in advance, preventing any disruption to the ordering process. This is one of the stages where ConChain reduces the possibility of unexpected failures at the orderer level, which will eventually lead to a higher number of successful transactions per second. After the validation, the dependency manager updates the structure of the transaction with two new extra properties ($readWallets$ and $writeWallets$) and then sends it for ordering. 

\begin{figure*}[!h]
\centering
\includegraphics[width=0.93\linewidth]{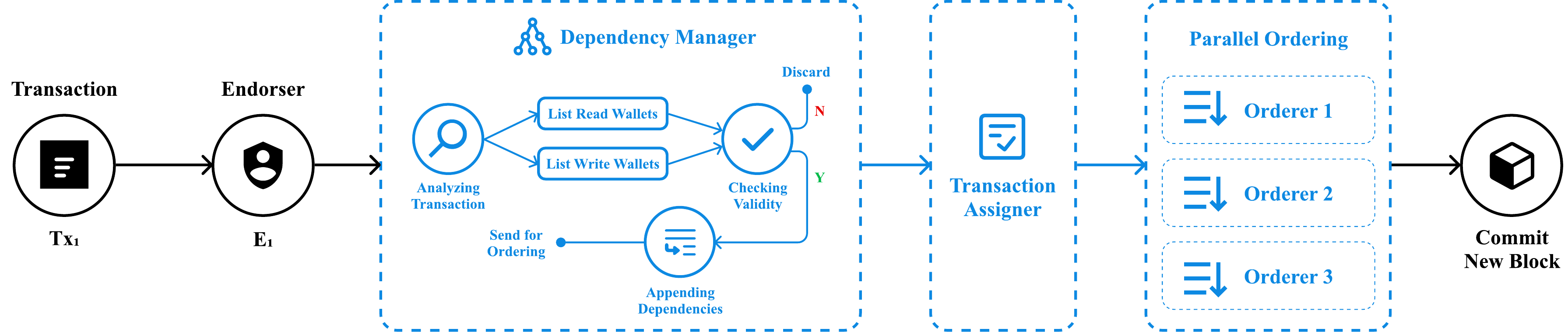}
        \caption{Architecture of our proposed approach (ConChain) and the workflow of Dependency Manager} \label{fig:dep}
\end{figure*}

Figure \ref{fig:dep} illustrates the entire ConChain architecture along with the workflow inside the dependency manager. Algorithm \ref{alg:dep} provides a logical representation of the dependency manager's workflow. For simplicity, we have omitted Hyperledger Fabric terminologies in the pseudocode. As described before, the dependency manager begins by checking the address type of each address in the transaction (Lines 1-3). Later, it validates each wallet (Lines 4-11) and discards transactions with invalid wallets (Line 11). For each valid wallet, it determines the access type and includes them in read/write groups (Lines 6-9). Finally, the dependency manager appends the dependent wallets to the transaction and returns the updated structure (Lines 12-14) for ordering.

\IncMargin{1em}
\setlength{\textfloatsep}{0pt}
 \begin{algorithm}
  \caption{\textbf{\texttt{AnalyzeDependency()}}} 
  \SetKwFunction{BuildTree}{AnalyzeDependency}
  \label{alg:dep}
  \SetKwInOut{Input}{Input}
  \SetKwInOut{Output}{Output}
  \Indm 
    \Input{Transaction $T_{1}$} 
    \Output{Update $T_{1}$ with dependencies}
    \Indp
    \For{each $addr$ in $T_1$}{
        \If{$addr.type$ == $WALLET$}{
            $wallets.\texttt{push}(addr)$
        }
    }
    
    \For{each $wallet$ in $wallets$}{
        \eIf{$\texttt{isValid}(wallet)$}{
            \If{$wallet.accessType$ == $WRITE$} {
                $writeWallets.\texttt{push}(wallet)$
            }
            \If{$wallet.accessType$ == $READ$} {
                $readWallets.\texttt{push}(wallet)$
            }
        }{
            $\texttt{discard}(T_1)$
        }
    }

    $T_1.readWallets \leftarrow readWallets$\\
    $T_1.writeWallets \leftarrow writeWallets$\\

    return $T_1$
\end{algorithm}
\DecMargin{1em}
While the dependency manager plays a crucial role in the overall network, it introduces some additional overhead. Although it doesn't significantly reduce the throughput, it does add extra processing delay to the conventional Fabric Architecture. In our test results, after adding the dependency manager, the average throughput dropped 40\% compared to the default fabric. To address this, we have implemented a parallel ordering process to significantly enhance network speed. However, instead of opting for a standard parallel processing approach, we have introduced another component to the network known as the Transaction Assigner.

\subsection{Transaction Assigner and Parallel Processing}
The transaction assigner acts as an orchestrator between the dependency manager and the orderers. This component efficiently determines the sequence of ordering transactions to maximize throughput and minimize conflicts. To achieve this, it maintains a queue of pending transactions from the dependency manager, with each transaction providing easy access to its dependent wallets. This accessibility allows the transaction assigner to check the availability of dependent wallets and prioritize ordering accordingly. 

Algorithm \ref{alg:assignTx} illustrates the logical workflow of the transaction assigner, functioning as a background job that regularly checks the top element in the transaction queue. Firstly, it accesses the read and write wallets from the transaction (Lines 1-2). Then, for each readWallet, it checks if the wallet is available or locked. If there is any locked wallet, it stops processing the transaction and pushes it back to the queue (Lines 3-7). Similarly, it also checks the availability of the writeWallets and pushes the transaction back if there is any locked wallet (Lines 10-14). If there are no locked wallets, that means this transaction is ready to be processed now. It pops the transaction from the queue (Lines 8, 9, 15, 16) and starts the next step of processing. Once a transaction finishes processing it automatically unlocks the wallet before committing the new block. 

\begin{figure}[!h]
\centering
\includegraphics[width=0.8\linewidth]{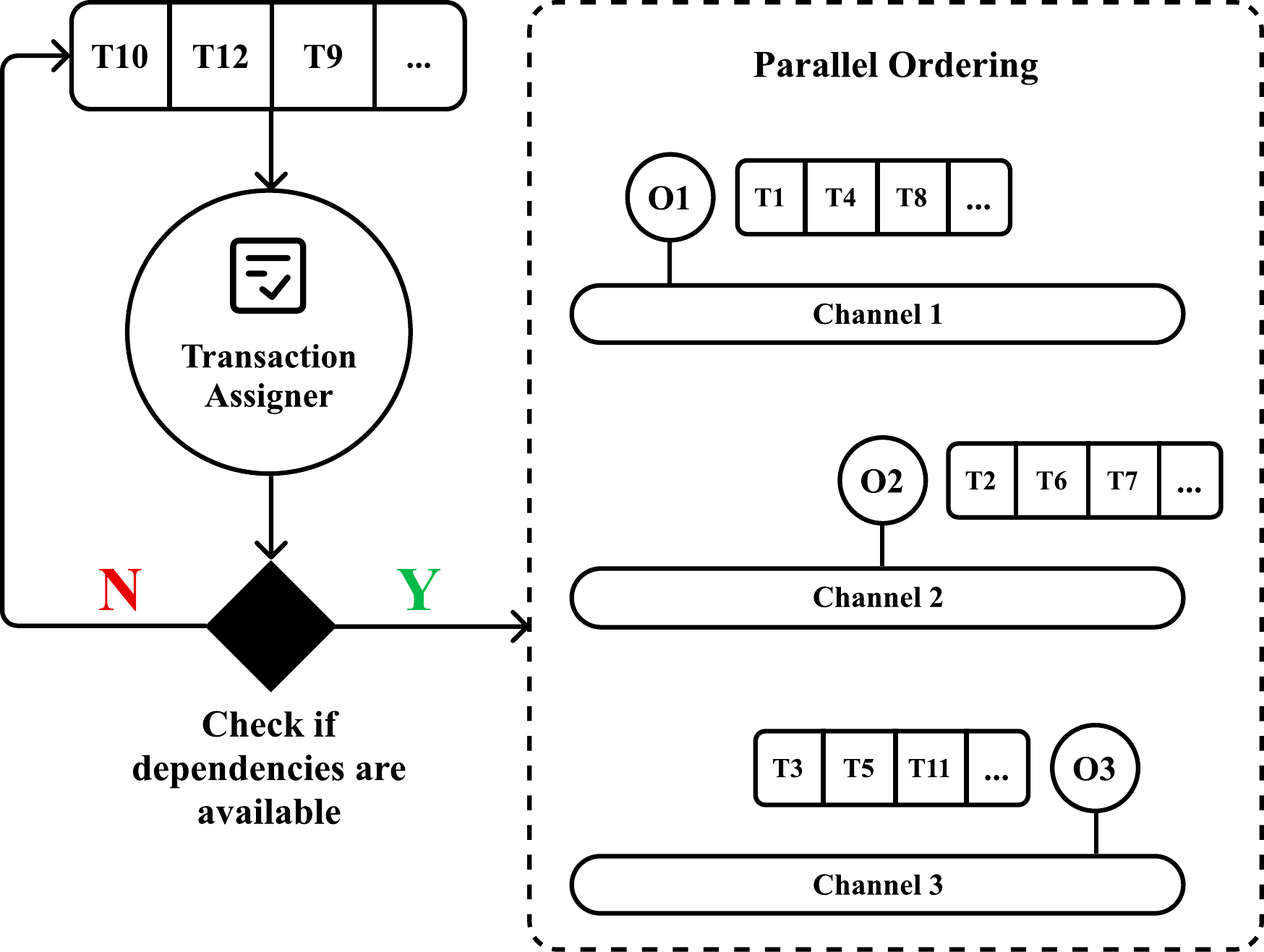}
        \caption{The structure and workflow of the Transaction Assigner and Parallel Ordering} \label{fig:par}
\end{figure}

Since we are implementing parallel ordering, we have created several channels with orderers for that purpose. Every channel is connected to all peers to ensure consistency across all channels during ordering. Figure \ref{fig:par} illustrates the workflow of the Transaction Assigner and Parallel Ordering. Each orderer has a queue limit. The transaction assigner checks if there is any orderer capable of processing another transaction (Line 17), sets that orderer's channel ID to the transaction's channel (Line 18), and finally assigns the transaction for ordering. Before assigning to the orderer, the dependent wallets will be locked (Lines 19-22). After the ordering is finished, the orderer will release the wallets by default nature.

In this way, our approach minimizes the possibility of conflicts. Additionally, priority-based ordering saves a significant amount of resources and waiting time by fully utilizing all the orderers' potential.

\IncMargin{1em}
\setlength{\textfloatsep}{0pt}
 \begin{algorithm}
  \caption{\textbf{\texttt{AssignTransaction()}}} 
  \SetKwFunction{BuildTree}{AssignTransaction}
  \label{alg:assignTx}
  \SetKwInOut{Input}{Input}
  \SetKwInOut{Output}{Output}
  \Indm 
    \Input{Transaction $T_{1}$} 
    \Output{Assign $T_{1}$ to a worker}
    \Indp
    $readWallets$ $\leftarrow$ \texttt{getReadRequests($T_{1}$)}\\
    $writeWallets$ $\leftarrow$ \texttt{getWriteRequests($T_{1}$)}\\
    \For{each $wallet$ in $readWallets$}{
        \If{\texttt{isLocked($wallet$)}} {
            \If{$queue$.\texttt{notInclude($T_1$)}}{
                $queue$.\texttt{push($T_1$)}\\
                \texttt{return}\\
            }
        }
    }

    \If{$queue$.\texttt{include($T_1$)}}{
        $queue$.\texttt{pop($T_1$)}
    }

    \For{each $wallet$ in $writeWallets$}{
        \If{\texttt{isLocked($wallet$)}} {
            \If{$queue$.\texttt{notInclude($T_1$)}}{
                $queue$.\texttt{push($T_1$)}\\
                \texttt{return}\\
            }
        }
    }

    \If{$queue$.\texttt{include($T_1$)}}{
        $queue$.\texttt{pop($T_1$)}\\
    }

    $c$ $\leftarrow$ \texttt{getAvailableChannel()}

    $T_1.\texttt{setChannel}(c)$

    \If{$c.orderer$ is not null}{
        \texttt{lock($readWallets$)}\\
	   \texttt{lock($writeWallets$)}\\
	   $c.orderer$.\texttt{assign($T_1$)}\\
    }
\end{algorithm}
\DecMargin{1em}

\section{Performance Analysis}
\label{sec:performance}
To evaluate the performance of our proposed approach, we utilized the same simulation environment detailed in Section \ref{sec:simulation}. In this iteration, we increased the number of nodes from 4 to 8 and simulated 100,000 randomly generated transactions for each of the six types, totaling 600,000 transactions. Subsequently, we calculated the average for each metric derived from these transactions.

Additionally, for performance comparison, we also simulated FastFabric \cite{gorenflo2020fastfabric}, which exhibited significant performance improvement over regular fabric. The overall simulation performance was measured using Hyperledger Caliper. This section will present a comparative analysis of the performance of our proposed approach.

\subsection{Success Rate}
\begin{figure*}
  \centering
  \begin{subfigure}[b]{0.24\textwidth}
  \centering
    \includegraphics[height=1.6in]{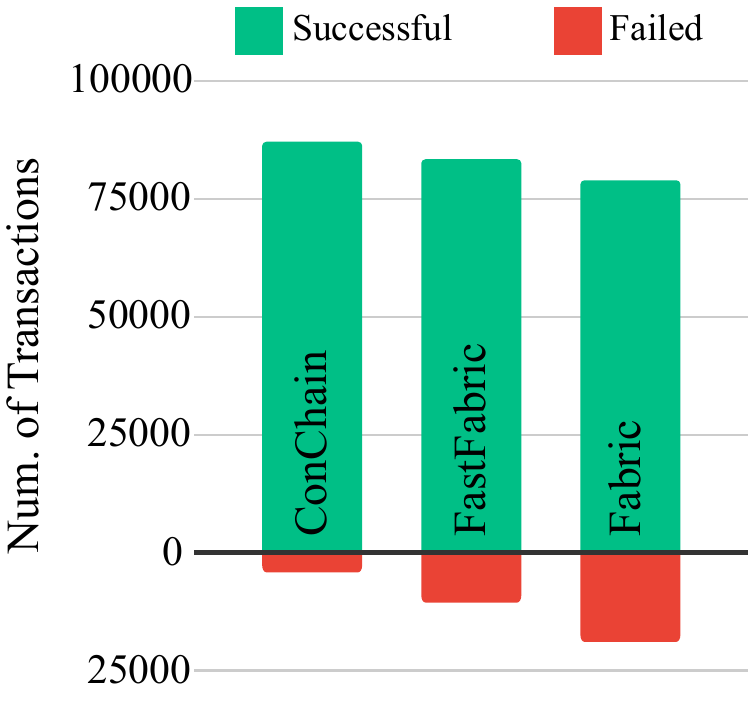}
            \caption{Success Rate Comparison} \label{fig:succ-rate}
    \end{subfigure}
  \begin{subfigure}[b]{0.24\textwidth}
  \centering
    \includegraphics[height=1.6in]{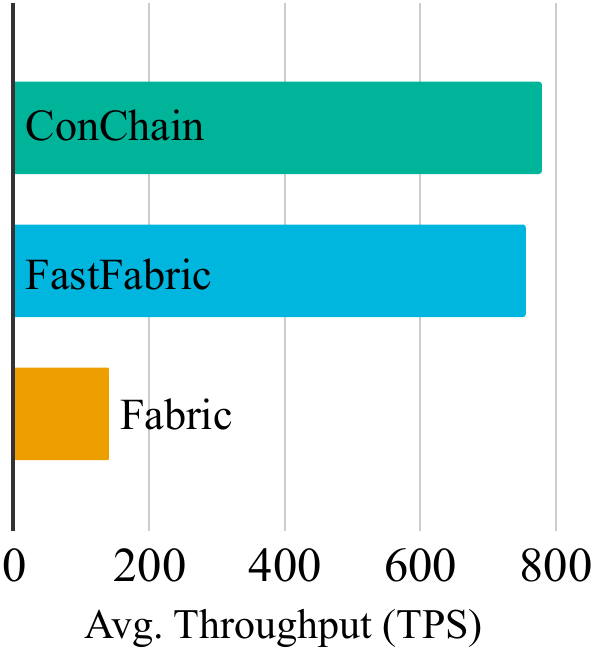}
        \caption{Throughput Comparison} \label{fig:tps}
    \end{subfigure}
    \begin{subfigure}[b]{0.24\textwidth}
    \centering
    \includegraphics[height=1.6in]{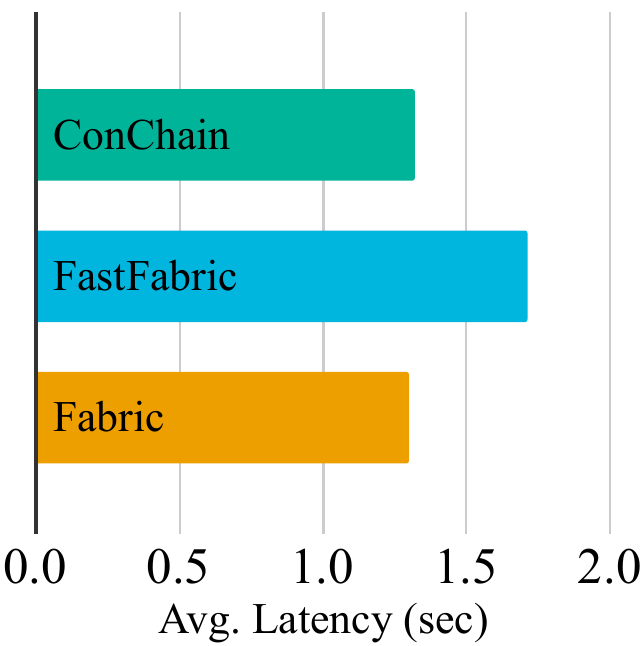}
            \caption{Latency Comparison} \label{fig:latency}
    \end{subfigure}
    \begin{subfigure}[b]{0.24\textwidth}
    \centering
    \includegraphics[height=1.6in]{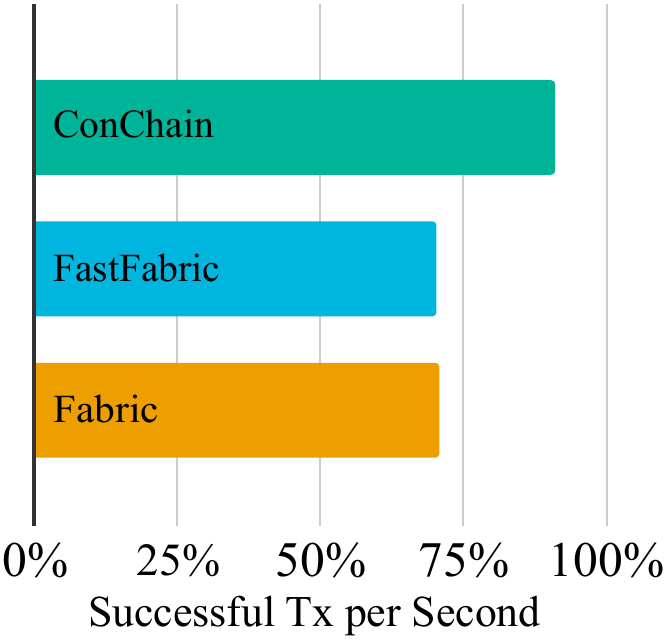}
        \caption{Successful TPS Comparison} \label{fig:stps}
    \end{subfigure}
    \caption{Performance Comparison with FastFabric and Default Fabric}
\end{figure*}

First of all, we measured the number of successful and failed transactions to conduct a comparison of success rates. As depicted in Figure \ref{fig:succ-rate}, it is evident that ConChain's success rate is significantly higher than the other two schemes. On average, ConChain achieves a 94\% success rate for all types of transactions. In comparison, FastFabric also demonstrates a respectable number of successful transactions, averaging 87\% when compared to regular Fabric. However, even though the success rate for Fabric improved with an increased number of transactions from the initial simulation, our approach consistently outperforms both in terms of the success rate of transactions.

\subsection{Latency}
Moving on to the critical metric of average latency, Figure \ref{fig:latency} illustrates that ConChain and Fabric exhibit similar latency, approximately 1.2 seconds. However, FastFabric shows slightly more latency than the others. Our System Architecture section highlighted that our Transaction Assigner optimally utilizes all orderers' potential. Consequently, despite the additional overhead from other modules, our approach maintains minimal latency, delivering performance comparable to regular fabric.

\subsection{Throughput}
As we focus on maximizing performance, throughput becomes a pivotal metric. The core innovation of ConChain is to minimize resource waste due to failures and maximize throughput. Figure \ref{fig:tps} reveals that ConChain outperforms the other two schemes. While FastFabric is primarily designed for higher throughput and shows over 6 times higher throughput compared to regular Fabric due to its optimized setup, ConChain slightly outperforms FastFabric in terms of throughput. This underscores that handling conflicts can yield significantly better performance using existing system components.

\subsection{Successful Transactions Per Second}
To demonstrate the most impactful contribution of ConChain, we measured the number of successful transactions per second. Given our approach's primary focus on minimizing failures due to conflicts, it excels in performance in this aspect as shown in Figure \ref{fig:stps}. In contrast to FastFabric and regular Fabric, ConChain processes significantly more successful transactions per second. While others exhibit around 60-70\% successful transactions, ConChain consistently achieves over 80\% success per second, with an average successful TPS of 86.3\%.

Our proposed approach's performance in comparison to alternative schemes highlights its consistent superiority across essential metrics. ConChain emerges as a standout performer, surpassing its counterparts in success rates, latency, throughput, and successful transactions per second. These results underscore the effectiveness of ConChain in optimizing system components and handling conflicts, positioning it as a promising solution for enhancing overall transaction performance.

\section{Conclusion}
\label{sec:conc}
In this study, we have provided a comprehensive exploration of the impact of contention on blockchain performance. The explanation of how contention affects the complexities of blockchain operations emphasizes the critical need to prioritize successful transactions per second above a purely quantitative focus on transactions per second. The novel approach that we proposed in this work, incorporating a dependency manager and parallel ordering, exhibited notably superior performance when compared to alternative approaches such as FastFabric and the default Fabric. We believe the implications of this improved performance are substantial, particularly in fostering the widespread adoption of blockchain technology. The ability to minimize conflicting transactions enhances the overall efficacy of blockchain networks, contributing to a more reliable and efficient system. As we look towards the future, our commitment involves making this solution pluggable, ensuring adaptability across diverse blockchain implementations. We aim to extend our focus to both public and private blockchains, further advancing the scalability and success of blockchain technology in various contexts.

\bibliographystyle{IEEEtran}
\bibliography{IEEEabrv,references}

\begin{thebibliography}{10}
\providecommand{\url}[1]{#1}
\csname url@samestyle\endcsname
\providecommand{\newblock}{\relax}
\providecommand{\bibinfo}[2]{#2}
\providecommand{\BIBentrySTDinterwordspacing}{\spaceskip=0pt\relax}
\providecommand{\BIBentryALTinterwordstretchfactor}{4}
\providecommand{\BIBentryALTinterwordspacing}{\spaceskip=\fontdimen2\font plus
\BIBentryALTinterwordstretchfactor\fontdimen3\font minus \fontdimen4\font\relax}
\providecommand{\BIBforeignlanguage}[2]{{%
\expandafter\ifx\csname l@#1\endcsname\relax
\typeout{** WARNING: IEEEtran.bst: No hyphenation pattern has been}%
\typeout{** loaded for the language `#1'. Using the pattern for}%
\typeout{** the default language instead.}%
\else
\language=\csname l@#1\endcsname
\fi
#2}}
\providecommand{\BIBdecl}{\relax}
\BIBdecl

\bibitem{HyperledgerFabric}
\BIBentryALTinterwordspacing
``Hyperledger fabric.'' [Online]. Available: \url{https://www.hyperledger.org /use/fabric}
\BIBentrySTDinterwordspacing

\bibitem{khan2018fast}
N.~Khan, ``Fast: a mapreduce consensus for high performance blockchains,'' in \emph{Proceedings of the 1st Workshop on Blockchain-enabled Networked Sensor Systems}, 2018, pp. 1--6.

\bibitem{gorenflo2020fastfabric}
C.~Gorenflo, S.~Lee, L.~Golab, and S.~Keshav, ``Fastfabric: Scaling hyperledger fabric to 20 000 transactions per second,'' \emph{International Journal of Network Management}, vol.~30, no.~5, p. e2099, 2020.

\bibitem{gorenflo2020xox}
C.~Gorenflo, L.~Golab, and S.~Keshav, ``Xox fabric: A hybrid approach to blockchain transaction execution,'' in \emph{2020 IEEE International Conference on Blockchain and Cryptocurrency (ICBC)}.\hskip 1em plus 0.5em minus 0.4em\relax IEEE, 2020, pp. 1--9.

\bibitem{amiri2019parblockchain}
M.~J. Amiri, D.~Agrawal, and A.~El~Abbadi, ``Parblockchain: Leveraging transaction parallelism in permissioned blockchain systems,'' in \emph{2019 IEEE 39th International Conference on Distributed Computing Systems (ICDCS)}.\hskip 1em plus 0.5em minus 0.4em\relax IEEE, 2019, pp. 1337--1347.

\bibitem{gervais2016security}
A.~Gervais, G.~O. Karame, K.~W{\"u}st, V.~Glykantzis, H.~Ritzdorf, and S.~Capkun, ``On the security and performance of proof of work blockchains,'' in \emph{Proceedings of the 2016 ACM SIGSAC conference on computer and communications security}, 2016, pp. 3--16.

\bibitem{larimer2013transactions}
D.~Larimer, ``Transactions as proof-of-stake,'' \emph{Nov-2013}, vol. 909, 2013.

\bibitem{xu9317791}
X.~Xu, X.~Wang, Z.~Li, H.~Yu, G.~Sun, S.~Maharjan, and Y.~Zhang, ``Mitigating conflicting transactions in hyperledger fabric-permissioned blockchain for delay-sensitive iot applications,'' \emph{IEEE Internet of Things Journal}, vol.~8, no.~13, pp. 10\,596--10\,607, 2021.

\bibitem{peng2022neuchain}
Z.~Peng, Y.~Zhang, Q.~Xu, H.~Liu, Y.~Gao, X.~Li, and G.~Yu, ``Neuchain: a fast permissioned blockchain system with deterministic ordering,'' \emph{Proceedings of the VLDB Endowment}, vol.~15, no.~11, pp. 2585--2598, 2022.

\bibitem{salehi2014taxonomy}
M.~A. Salehi, J.~H. Abawajy, and R.~Buyya, ``Taxonomy of contention management in interconnected distributed systems.'' 2014.

\bibitem{salehi2014contention}
M.~A. Salehi, A.~N. Toosi, and R.~Buyya, ``Contention management in federated virtualized distributed systems: implementation and evaluation,'' \emph{Software: Practice and Experience}, vol.~44, no.~3, pp. 353--368, 2014.

\bibitem{kostin2000randomized}
A.~E. Kostin, I.~Aybay, and G.~Oz, ``A randomized contention-based load-balancing protocol for a distributed multiserver queuing system,'' \emph{IEEE Transactions on Parallel and Distributed Systems}, vol.~11, no.~12, pp. 1252--1273, 2000.

\bibitem{zhang2009location}
B.~Zhang and B.~Ravindran, ``Location-aware cache-coherence protocols for distributed transactional contention management in metric-space networks,'' in \emph{2009 28th IEEE International Symposium on Reliable Distributed Systems}.\hskip 1em plus 0.5em minus 0.4em\relax IEEE, 2009, pp. 268--277.

\bibitem{fabricCRDT}
\BIBentryALTinterwordspacing
P.~Nasirifard, R.~Mayer, and H.-A. Jacobsen, ``Fabriccrdt: A conflict-free replicated datatypes approach to permissioned blockchains,'' in \emph{Proceedings of the 20th International Middleware Conference}, ser. Middleware '19.\hskip 1em plus 0.5em minus 0.4em\relax New York, NY, USA: Association for Computing Machinery, 2019, p. 110–122. [Online]. Available: \url{https://doi.org/10.1145/3361525.3361540}
\BIBentrySTDinterwordspacing

\bibitem{xuLocking}
\BIBentryALTinterwordspacing
L.~Xu, W.~Chen, Z.~Li, J.~Xu, A.~Liu, and L.~Zhao, ``Locking mechanism for concurrency conflicts on hyperledger fabric,'' in \emph{Web Information Systems Engineering – WISE 2019: 20th International Conference, Hong Kong, China, January 19–22, 2020, Proceedings}.\hskip 1em plus 0.5em minus 0.4em\relax Berlin, Heidelberg: Springer-Verlag, 2020, p. 32–47. [Online]. Available: \url{https://doi.org/10.1007/978-3-030-34223-4\_3}
\BIBentrySTDinterwordspacing

\bibitem{kelkar2020order}
M.~Kelkar, F.~Zhang, S.~Goldfeder, and A.~Juels, ``Order-fairness for byzantine consensus,'' in \emph{Advances in Cryptology--CRYPTO 2020: 40th Annual International Cryptology Conference, CRYPTO 2020, Santa Barbara, CA, USA, August 17--21, 2020, Proceedings, Part III 40}.\hskip 1em plus 0.5em minus 0.4em\relax Springer, 2020, pp. 451--480.

\bibitem{helix}
D.~Yakira, A.~Asayag, G.~Cohen, I.~Grayevsky, M.~Leshkowitz, O.~Rottenstreich, and R.~Tamari, ``Helix: A fair blockchain consensus protocol resistant to ordering manipulation,'' \emph{IEEE Transactions on Network and Service Management}, vol.~18, no.~2, pp. 1584--1597, 2021.

\bibitem{goel2018resource}
S.~Goel, A.~Singh, R.~Garg, M.~Verma, and P.~Jayachandran, ``Resource fairness and prioritization of transactions in permissioned blockchain systems (industry track),'' in \emph{Proceedings of the 19th International Middleware Conference Industry}, 2018, pp. 46--53.

\bibitem{SmallBank}
\BIBentryALTinterwordspacing
``Smallbank benchmark.'' [Online]. Available: \url{https://hstore.cs.brown.edu/ documentation/deployment/benchmarks/smallbank}
\BIBentrySTDinterwordspacing

\bibitem{benoit2021diablo}
H.~Benoit, V.~Gramoli, R.~Guerraoui, and C.~Natoli, ``Diablo: A distributed analytical blockchain benchmark framework focusing on real-world workloads,'' 2021.

\bibitem{HyperledgerCaliper}
\BIBentryALTinterwordspacing
``Hyperledger caliper.'' [Online]. Available: \url{https://www.hyperledger.org /use/caliper}
\BIBentrySTDinterwordspacing

\end{thebibliography}

\end{document}